\documentclass[12pt]{amsart}

\usepackage{amssymb}
\usepackage{amsmath}
\usepackage{amsfonts}
\usepackage{graphicx}
\usepackage{epstopdf}
\usepackage{caption}
\title{Quadratic approximation of slow factor of volatility
in a Multi-factor Stochastic volatility Model}
\author{Gifty Malhotra}
\email{giftymalhotra@dtu.ac.in}
\author{R. Srivastava}
\email{rsrivastava@dce.ac.in}
\author{H.C. Taneja}
\email{hctaneja@dce.ac.in}
\address{Department of Applied Mathematics,	
Delhi Technological University,
Delhi(India)-110042}

\begin{document}

\begin{abstract}
In the present work, we propose a new multifactor stochastic volatility model in which slow factor of volatility is approximated by a parabolic arc. We retain ourselves to the perturbation technique to obtain approximate expression for European option prices. We introduce the notion of modified Black-Scholes price. We obtain a simplified expression for European option price which is perturbed around the modified Black-Scholes price and have also obtained the expression of modified price in terms of Black-Scholes price.
\smallskip


\end{abstract}
\maketitle

 \section{Introduction}
       Stochastic volatility models are prominent in option valuation literature as they are able to coalesce many stylized facts about volatility namely volatility smile, mean reversion, volatility clustering etc.(eg. see Bakshi, Cao and Chen $(1997)$, Bates $(2000)$, Chernov and Ghysels $(2000)$, Gatheral $(2006)$etc.). Single factor stochastic volatility models are being discussed in many papers(eg. see Hull and white $(1987)$, Stein-Stein $(1991)$, Heston $(1993)$, Ball-Roma $(1994)$ etc.). Volatility smile can be generated by single factor stochastic volatility models but its time varying nature remains unexplained by these models.\\
       Christoffersen et al. $(2009)$ strongly argued the need of multifactor stochastic volatility models to capture some of the most salient stylized facts in index options. They demonstrated the role of multiple factors in capturing term structure and moneyness effects.\\
       Multifactor stochastic volatility models are much recent and very significant in option valuation literature. Alizadeh et al. $(2002)$ found the evidence of two factors in volatility with one highly persistent factor and other quickly mean reverting factor. Extending this idea, Fouque et al. $(2003a)$ proposed a two factor stochastic volatility model with one fast mean reverting factor and another slowly varying factor. They have showed that slow varying volatility factor is essential for options with longer maturity and used perturbation analysis in context of pricing options. For more details one can refer Fouque et al. $(2011)$. As persistence of volatility should be given importance and must be incorporated in any volatility model (Robert F Engle et al. $(2001)$) and knowing the fact that multifactor stochastic volatility models give better results for the options with medium-longer maturity (Fatone et al. $(2009)$), the slow factor of volatility, which is highly persistent, is important and its dynamics can not be ignored. Also in perturbation analysis given by Fouque et al. $(2003a)$ the approximate option price depends upon slow factor of volatility but independent of fast factor.\\
       We propose a multifactor model to give importance to slow factor of volatility by taking it mean reverting and approximating it by a parabolic arc. Also, with this model we have derived the pricing formula for European call options. We have also introduced the notion of modified Black Scholes operator.
       The paper is organised as follows: In Section $2$ we introduce the multifactor stochastic volatility model. Option pricing equation  and asymptotic expansion of price is discussed in Section $3$ and $4$ respectively. Approximate price of European option is given in Section $5$ and Section $6$ includes the conclusion.
\section{Model under consideration}
Let $X_t$ be the price of underlying asset (non dividend paying), $P^{*}$ be the risk neutral probability measure and r be the risk free rate of interest. Under $P^{*}$, the dynamics of $X_t$ be given by a diffusion process as:
\begin{equation}
dX_t = rX_t dt + \sigma X_t dW_{t}^{x}
 \end{equation}
 Here,
 \begin{equation}
 \sigma = f(Y_t, Z_t)
 \end{equation}
 is stochastic volatility driven by two factors $Y_t$ and $Z_t$ which are respectively the fast scale and slow scale factors of volatility. $W_{t}^{x}$ is standard Brownian motion. We consider the dynamics of fast volatility factor $Y_t$ as given in Fouque et al. $(2003a)$:
 \begin{equation}
dY_t = \frac{1}{\epsilon}(m - Y_t) dt + \frac{\nu\sqrt {2}}{\sqrt{\epsilon}} dW_{t}^{y}
\end{equation}
which is an Ornstein-Uhlenbeck (OU) process with long run distribution $N(m,\nu^{2})$ and is reverting on the short time scale $\epsilon$ around its long run mean value $m$ with $1/\epsilon$ as its rate of mean reversion. Its volatility of volatility (vol-vol) parameter is $\frac{\nu\sqrt {2}}{\sqrt{\epsilon}}$. $W_{t}^{y}$ is standard Brownian motion.\\ $W_{t}^{x}$ and $W_{t}^{y}$ have the correlation structure:

  $$E[dW_{t}^{x}.dW_{t}^{y}] = \rho_{xy} dt$$

  Where, $\rho_{xy}$ represents the correlation between two standard Brownian motions.
  Slow factor of volatility is persistent and empirically it is mean reverting too. But mean reversion can be clearly observed for long term options. We take quadratic arc approximation to capture slow volatility factor given by:
\begin{equation}
Z_t= \mathcal{A}t^{2}+\mathcal{B}t+\mathcal{C}+\alpha_{t}
\end{equation}
where $\alpha_{t}$ represents the error term in the approximation of $Z_{t}$ with $\mathcal{A}\neq 0$.
We can justify this approximation of persistent factor of volatility, $Z_t$ from $S\&P 500$ historic volatility data given in
\begin{figure}[!htb]
\centering
\vspace*{-2.5cm}
\includegraphics[width=8cm]{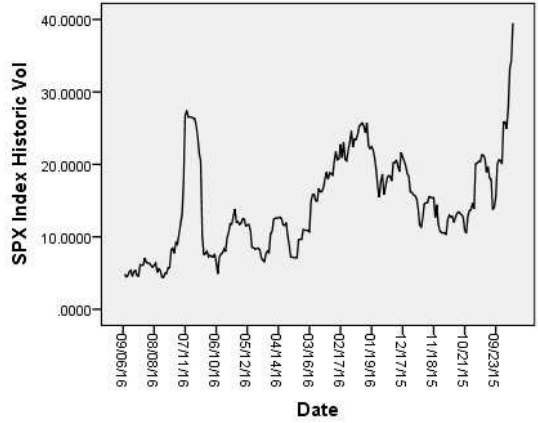}
\caption{Historic volatility of SPX index from Septem \hspace*{2cm}-ber $8,2015$  to September $6,2016$, clearly mean \hspace*{2cm}reverting about 20$\%$(say)}
\label{hisvol}
\vspace*{-0.7cm}
\end{figure}
 fig.\ref{hisvol}\\

We are also justifying this choice of approximation by considering slow volatility factor $Z_t$ to follow a diffusion process:
\begin{equation}
dZ_t = k(m^{'} - Z_t) dt + \eta dW_{t}^{z}
\end{equation}
where $W_{t}^{z}$ is also a standard Brownian motion having the correlation structure with $W_{t}^{x}$ and $W_{t}^{y}$ as :

 $$E[dW_{t}^{x}.dW_{t}^{z}] =  \rho_{xz} dt$$
  $$E[dW_{t}^{y}.dW_{t}^{z}]= \rho_{yz}dt$$
 where the correlation coefficients $\rho_{xz}$ and $\rho_{yz}$ are such that $\rho_{xy}^{2} < 1 , \rho_{xz}^{2} < 1 , \rho_{yz}^{2} < 1$  and $1 + 2\rho_{xy}\rho_{xz}\rho_{yz}-\rho_{xy}^{2}-\rho_{xz}^{2}-\rho_{yz}^{2} > 0$ for the positive definiteness of the covariance matrix of three Brownian motions.\\ Here, $Z_t$ follows Ornstein-Uhlenbeck (OU)process with long run distribution $N(m^{'},\frac{\eta^{2}}{2k})$ and is reverting on long time scale $1/k$ around its long run mean value $m^{'}$. The rate of mean reversion for $Z_t$ is $k$ and its vol-vol parameter is $\eta$.
\\ On solving, $(5)$ becomes:
\begin{equation}
Z_t=\biggl[\frac{(Z_{0}-m^{'})}{2}k^2\biggr]t^2+[-k(Z_{0}-m^{'})]t+Z_{0}+\delta_{t}+\beta_{t}
\end{equation}
comparing it with $(4)$, we obtain
$$\mathcal{A}=\frac{(Z_{0}-m^{'})}{2}k^{2},$$

$$\mathcal{B}=-(Z_0-m^{'})k,$$

$$\mathcal{C}=Z_{0},$$

and
$$\alpha_{t}=\delta_{t}+\beta_{t}$$

 Here,$Z_{0}$ represents the initial value of slow factor of volatility. To assure that $\mathcal{A}\neq 0$ we assume that $Z_0\neq m^{'}$. $\delta_{t}$ is the truncation error and $\beta_{t}$ involving $\eta$, is randomness in the value of slow factor of volatility $Z_t$.\\ We are assuming here that the error term $\alpha_{t}$ is negligible as truncation error can be neglected and vol-vol parameter $\eta$ of slow factor of volatility can be neglected under parabolic approximation.

The approximated value of $Z_t$ is a function of t. also,
\begin{equation}
\frac{\partial}{\partial z}=\biggl(\frac{1}{2\mathcal{A}t+\mathcal{B}+\zeta_t}\biggr)\frac{\partial}{\partial t}
\end{equation}
and
\begin{equation}
\frac{\partial^{2}}{\partial z^{2}}=\frac{1}{(2\mathcal{A}t+\mathcal{B}+\zeta_t)^{2}}\biggl[\frac{\partial^{2}}{\partial t^{2}}-\biggl(\frac{2\mathcal{A}+\zeta_{t}^{'}}{2\mathcal{A}t+\mathcal{B}+\zeta_{t}}\biggr)\frac{\partial}{\partial t}\biggr]
\end{equation}
where,

$$\zeta_t=\frac{\partial \alpha_{t}}{\partial t}$$.

These expressions will be needed for pricing in the upcoming sections. So in the nutshell, our multifactor stochastic volatility model under risk neutral probability measure $P^{*}$, which is already chosen by the market is:
\begin{equation}
dX_t = rX_t dt + \sigma X_t dW_{t}^{x}
\end{equation}
$$dY_t = \frac{1}{\epsilon}(m - Y_t) dt + \frac{\nu\sqrt {2}}{\sqrt{\epsilon}} dW_{t}^{y}$$
$$Z_t= \mathcal{A}t^{2}+\mathcal{B}t+\mathcal{C}+\alpha_{t}$$
where, $$ \mathcal{A}=\frac{(Z_{0}-m^{'})}{2}k^{2},$$
$$ \mathcal{B}=-(Z_0-m^{'})k,$$
$$ \mathcal{C}=Z_{0}$$

 which we have obtained by considering diffusion process
 $$dZ_t = k(m^{'} - Z_t) dt + \eta dW_{t}^{z}$$ for $Z_{t}$. Next we'll write the pricing equation for European options.

\section{Pricing Equation}
The price of European call option with payoff function $h(X_T)$ under the risk neutral probability measure $P^{*}$ is conditional expectation of discounted payoff given as:
\begin{equation}
P^{\epsilon}(t,x,y,z)= E^{*}\{ e^{-r(T-t)}h(X_T)|X_t=x,Y_t=y,Z_t=z\},
\end{equation}
By Feynman-Kac formula, $P^{\epsilon}(t,x,y,z)$ satisfies the following parabolic PDE:
\begin{equation}
\mathcal{L}^{\epsilon}P^{\epsilon}(t,x,y,z)=0,
\end{equation}
with the boundary condition
$$P^{\epsilon}(T,X_T,Y_T,Z_T)=h(X_T),$$
where the operator $\mathcal{L}^{\epsilon}$ is given by:
\begin{equation}
\mathcal{L}^{\epsilon}=\frac{1}{\epsilon}\mathcal{L}_{0}+\frac{1}{\sqrt\epsilon}\mathcal{L}_{1}+\mathcal{L}_{2},
\end{equation}
with
\begin{equation}
\mathcal{L}_{0}=(m-y)\frac{\partial}{\partial{y}}+\nu^{2}\frac{\partial^{2}}{\partial{y^2}},
\end{equation}
\begin{equation}
\mathcal{L}_{1}=\rho_{xy}\nu\sqrt{2}f(y,z) x\frac{\partial^{2}}{\partial x \partial y}+\rho_{yz}\nu\sqrt{2}\eta\frac{\partial^{2}}{\partial y\partial z},
\end{equation}
and
\begin{equation}
\mathcal{L}_{2}=\frac{\partial}{\partial{t}}+\frac{1}{2}f^{2}(y,z)x^{2}\frac{\partial^{2}}{\partial x^{2}}+r(x\frac{\partial}{\partial x}- .)+\rho_{xz}\eta f(y,z) x \frac{\partial^{2}}{\partial x\partial z}+\frac{1}{2}\eta^{2}\frac{\partial^{2}}{\partial z^{2}}+k(m^{'}-z)\frac{\partial}{\partial z}.
\end{equation}
On putting the value of $\frac{\partial}{\partial z}$ and $\frac{\partial^{2}}{\partial z^{2}}$ from $(7)$ and $(8)$ in $(14)$ and $(15)$ and on solving, we get
\begin{equation}
\mathcal{L}_{1}=\nu\sqrt{2}\biggl[\rho_{xy}f(y,z) x\frac{\partial^{2}}{\partial x \partial y}+\frac{\rho_{yz}}{2\mathcal{A}t+\mathcal{B}+\zeta_t}\eta\frac{\partial^{2}}{\partial y\partial t}\biggr],
\end{equation}
and
\begin{align}
\mathcal{L}_{2}=& \biggl[1+\frac{k(m^{'}-z)}{2\mathcal{A}t+\mathcal{B}+\zeta_t}-
\frac{1}{2}\eta^{2}\frac{2\mathcal{A}+\zeta_{t}^{'}}{(2\mathcal{A}t+\mathcal{B}+\zeta_{t})^{3}}\biggr]\frac{\partial}{\partial{t}}+\frac{1}{2}f^{2}(y,z)x^{2}\frac{\partial^{2}}{\partial x^{2}}+r(x\frac{\partial}{\partial x}- .)\\&\nonumber+\biggl(\frac{\rho_{xz}\eta f(y,z)x}{2\mathcal{A}t+\mathcal{B}+\zeta_t}\biggr) \frac{\partial^{2}}{\partial x\partial t}+\frac{1}{2}\eta^{2}\frac{1}{(2\mathcal{A}t+\mathcal{B}+\zeta_t)^{2}}\frac{\partial^{2}}{\partial t^{2}}
\end{align}
where, $z=Z_{t}$ is given by $(6)$\\
From $(11)$ and $(12)$ we have,
\begin{equation}
(\frac{1}{\epsilon}\mathcal{L}_{0}+\frac{1}{\sqrt\epsilon}\mathcal{L}_{1}+\mathcal{L}_{2})P^{\epsilon}(t,x,y,z)=0
\end{equation}
This is the required pricing equation where $P^{\epsilon}$ is the solution of above parabolic PDE.
\section{Asymptotic Expansion}
For asymptotic expansion, expand $P^{\epsilon}$ in powers of $\sqrt\epsilon$, i.e.
\begin{equation}
P^{\epsilon}= P_0 + \sqrt\epsilon P_{1} + \epsilon P_{2}+ ...
\end{equation}
Using it in $(18)$ gives
\begin{equation}
(\frac{1}{\epsilon}\mathcal{L}_{0}+\frac{1}{\sqrt\epsilon}\mathcal{L}_{1}+\mathcal{L}_{2})(P_0 + \sqrt\epsilon P_{1} + \epsilon P_{2}+ ...)=0
\end{equation}
Similar expansion is considered in Fouque et al. $(2003a,2003b)$ for single factor and for two factors of volatility respectively. We have considered the expansion in powers of $\sqrt\epsilon$ only because we have approximated the slow factor of volatility.\\
Terms of order $\frac{1}{\epsilon}$:

$$\mathcal{L}_{0}P_{0}=0$$

\begin{equation}
\Rightarrow P_{0}=P_{0}(t,x,z)
\end{equation}
which is independent of $y$ but depending on $z$, the slow factor of volatility.\\
Terms of order $\frac{1}{\sqrt\epsilon}$:

$$\mathcal{L}_{1}P_{0}+\mathcal{L}_{0}P_{1}=0$$

as $P_{0}$ is independent of $y$

$$\Rightarrow \mathcal{L}_{0}P_{1}=0$$

\begin{equation}
\Rightarrow P_{1}=P_{1}(t,x,z)
\end{equation}
which is again independent of $y$ but depending on $z$, the slow factor of volatility.\\
Terms of order $1$:

$$\mathcal{L}_{2}P_{0}+\mathcal{L}_{1}P_{1}+\mathcal{L}_{0}P_{2}=0$$

\begin{equation}
\Rightarrow \mathcal{L}_{0}P_{2}+\mathcal{L}_{2}P_{0}=0
\end{equation}
$\because P_{1}$ is independent of $y$
and it is Poisson equation in $P_2$ with respect to $y$ with Fredholm solvability condition:

$$E_{y}[\mathcal{L}_{2}P_{0}] = 0$$

\begin{equation}
\Rightarrow E_{y}[\mathcal{L}_{2}]P_{0} = 0
\end{equation}
where, $E_{y}[\mathcal{L}_2]$ is the average of $\mathcal L_2$ w.r.t $y$. For simplification, we neglect the error term involving truncation and randomness in $(6)$ assuming $\delta_{t}\rightarrow 0$ and vol-vol parameter of $Z_{t},  \eta\approx 0$. Therefore, $(17)$ is reduced to

$$\mathcal{L}_{2}=\biggl[1+\frac{1-kt+\frac{k^{2}t^{2}}{2}}{1-kt}\biggr]\frac{\partial}{\partial{t}}+\frac{1}{2}f^{2}(y,z)x^{2}\frac{\partial^{2}}{\partial x^{2}}+r(x\frac{\partial}{\partial x}- .)$$

take,

$$\frac{1-kt+\frac{k^{2}t^{2}}{2}}{1-kt}=\gamma$$

with $kt\neq 1$  and $\gamma\neq0$
\begin{equation}
\Rightarrow \mathcal{L}_{2}=[1+\gamma]\frac{\partial}{\partial{t}}+\frac{1}{2}f^{2}(y,z)x^{2}\frac{\partial^{2}}{\partial x^{2}}+r(x\frac{\partial}{\partial x}- .)
\end{equation}
and,
\begin{equation}
E_{y}[\mathcal{L}_{2}]=[1+\gamma]\frac{\partial}{\partial{t}}+\frac{1}{2}\overline \sigma^{2}(z)x^{2}\frac{\partial^{2}}{\partial x^{2}}+r(x\frac{\partial}{\partial x}- .)
\end{equation}
where,

$$\overline\sigma(z)=E_{y}[f(y,z)]$$

which is a function of slow factor of volatility. We call $E_{y}[\mathcal{L}_{2}]$ as $\gamma-$ modified Black-Scholes operator with volatility $\overline\sigma(z)$. So, in $(24)$, $P_{0}$ is modified Black-Scholes price. After some calculation, expression of $P_{0}$ in terms of Black-Scholes price is given as:
\begin{equation}
P_{0}(t,x,z)=\frac{|kt-2|^{\frac{a-2r}{k}}e^{\frac{2r-a}{k(|kt-2|)}}}{e^{\frac{(2r-a)t}{2}}}Q_{0}(t,x,\overline\sigma(z))
\end{equation}
where, $Q_{0}$ is classical Black-Schole price and $m$ is the constant s.t. $a \neq 2r$ for $0\leq t<T$. Equality holds at maturity to satisfy the boundary condition

$$P_{0}(T,X_{T},Z_{T})= h(X_{T})$$

To approximate the value of $a$, we consider the numerical value of $(kt-2)$ which can be calculated from options data. We consider S$\&$P $500$ index option data with option starting from January $4,2016$ maturing on June $30,2016$. For different moneyness,we got $a \approx 0.05$ \\
 Also, from $(23)$, $\mathcal{L}_{0}P_{2}=-\mathcal{L}_{2}P_{0}$
$$\Rightarrow \mathcal{L}_{0}P_{2}=-[\mathcal{L}_{2}P_{0}-E_{y}[\mathcal{L}_{2}]P_{0}]$$
\begin{equation}
\Rightarrow P_{2}=-\mathcal{L}_{0}^{-1}[\mathcal{L}_{2}-E_{y}[\mathcal{L}_{2}]]P_{0}
\end{equation}
Terms of order $\sqrt \epsilon$:

$$\mathcal{L}_{0}P_{3}+\mathcal{L}_{2}P_{1}+\mathcal{L}_{1}P_{2}=0$$

which is Poisson equation in $P_3$ with respect to $y$ with Fredholm solvability condition:

$$E_{y}[\mathcal{L}_{2}P_{1}+\mathcal{L}_{1}P_{2}]=0$$

Considering $P_{2}$ from $(28)$ and after solving, we get
\begin{equation}
E_{y}[\mathcal{L}_{2}]P_{1}=\mathcal{G}P_{0}
\end{equation}
where,

$$\mathcal{G}=E_{y}[\mathcal{L}_{1}\mathcal{L}_{0}^{-1}[\mathcal{L}_{2}-E_{y}[\mathcal{L}_{2}]]]$$

On solving,
\begin{equation}
\mathcal{G}= \frac{\nu\rho_{xy}}{\sqrt{2}}E_{y}[f\frac{\partial\phi}{\partial y}]x\frac{\partial}{\partial x} x^{2} \frac{\partial^{2}}{\partial x^{2}}
\end{equation}
For first order approximation, we need the expression for $P_{1}$ along with $P_{0}$. With some simplification, one can easily verify that the term
\begin{equation}
P_{1}(t,x,z)=2\biggl[\frac{1}{k}\log \biggl(\frac{kT-2}{kt-2}\biggr)+\frac{T-t}{(kT-2)(kt-2)}\biggr]\mathcal{G}P_{0}
\end{equation}
is the unique solution of $(29)$
with boundary condition

$$P_{1}(T,X_T,Z_T)=0$$

where $\mathcal{G}$ is given by $(30)$.
\section{Approximate Option Price}
From the above calculation we get the First order approximation of option price as
$$P^{\epsilon}\approx \hat P^{\epsilon}= P_{0}+\sqrt\epsilon P_{1}$$
i.e.
\begin{equation}
P^{\epsilon}\approx \hat P^{\epsilon}= P_{0}+\sqrt\epsilon\biggl(2\biggl[\frac{1}{k}\log \biggl(\frac{kT-2}{kt-2}\biggr)+\frac{T-t}{(kT-2)(kt-2)}\biggr]\mathcal{G}\biggr)P_{0}
\end{equation}
where $P_{0}$ is given by $(27)$\\
Now, $(32)$ can be written as:
\begin{equation}
 P^{\epsilon}\approx (1+g) \biggl[Q_{0}+\sqrt\epsilon\biggl(2\biggl[\frac{1}{k}\log \biggl(\frac{kT-2}{kt-2}\biggr)+\frac{T-t}{(kT-2)(kt-2)}\biggr]VD_{1}D_{2}\biggr)Q_{0}\biggr]
 \end{equation}
Here,
\begin{equation}
V=\frac{\nu\rho_{xy}}{\sqrt{2}}E_{y}[f\frac{\partial\phi}{\partial y}]
\end{equation}
and $\phi$ is the solution of
\begin{equation}
\mathcal L_{0}\phi_{y,z}=f^{2}(y,z)-\overline\sigma^{2}(z)
\end{equation}
$\mathcal L_{0}$ is given by $(13)$.\\
 Equation $(33)$ gives the required first order approximation of option price $P^{\epsilon}$
 where,
 \begin{equation}
1+g =  \frac{|kt-2|^{\frac{a-2r}{k}}e^{\frac{2r-a}{k(|kt-2|)}}}{e^{\frac{(2r-a)t}{2}}}
 \end{equation}
such that $g$ converges to zero at maturity.\\ We name $1+g$ as modification factor. We are interested in its value. Numerically, if we take $a=0.05, r=0.0264, k=0.008$, for $t=0,0.25$ and $0.5$, we get $1+g\approx0.934$. It will significantly improve the pricing. We have used the market data to obtain these parameters.

The accuracy of this first order approximation and calibration of effective stochastic volatility parameter $V$ can be done as already explained in Fouque et al. $(2003)$.

\section{Conclusion}
Slow factor of volatility is persistent and its dynamics can't be ignored. No doubt it is stochastic, but approximating it with a deterministic arc makes the calculation easy and give a simplified expression for the price of European call option. This price is perturbed around the modified Black-Scholes price. We have also given the expression of modified price in terms of Black-Scholes price and have calculated the modification factor using S$\&$P $500$ index options data.

\end{document}